\documentstyle[psfig,12pt,aps,prb,floats,tighten]{revtex}
\newcommand{\lcaption}[3]
  {\caption{{\hspace{-18truemm}\protect\parbox[t]{ #1 
    }{\setlength{\parindent}{18truemm} #3 }\hskip #2
    }}}
\begin{document}
\draft

\title{Enhancement of Coulomb interactions in semiconductor
nanostructures by dielectric confinement}

\author{\underline{G. Goldoni}$^{1,2}$, F. Rossi$^{1,3}$, A. Orlandi$^{1,2}$, 
M. Rontani$^{1,2}$, F. Manghi$^{1,2}$, E. Molinari$^{1,2}$}

\address{${}^{1}$ Istituto Nazionale per la Fisica della Materia (INFM)}
\address{${}^{2}$ Dipartimento di Fisica, Universit\`a di Modena, Via Campi
213/A, I-41100 Modena, Italy}
\address{${}^{3}$ Dipartimento di Fisica, Politecnico di Torino, C.so Duca 
degli
Abruzzi 24,  I-10129 Torino, Italy}
\date{\today}
\maketitle

\begin{abstract}
\baselineskip=13.5pt

We present a theoretical analysis of the effect of dielectric
confinement on the Coulomb interaction in dielectrically modulated
quantum structures. We discuss the implications of the strong
enhancement of the electron-hole and electron-electron coupling for
two specific examples: (i) GaAs-based quantum wires with remote oxide
barriers, where combined quantum and dielectric confinements are
predicted to lead to room temperature exciton binding, and (ii)
semiconductor quantum dots in colloidal environments, where the
many-body ground states and the addition spectra are predicted to be
drastically altered by the dielectric environment.

\end{abstract} 
\bigskip 

\baselineskip=13.5pt

When a semiconductor nanostructure is embedded in a medium with a
smaller dielectric constant, the Coulomb interaction between quantum
confined states may be enhanced by virtue of the polarization charges
which form at the dielectrically mismatched interfaces \cite{Keldysh}.
While this effect is relatively small and usually neglected in
conventional semiconductor heterostructures (e.g., GaAs/AlAs), we will
show that, for hybrid semiconductor nanostructures surrounded by an
organic or dielectric medium the enhancement can be large and must be
taken into account for a realistic description of Coulomb correlated
quantum states. Beside being quantitatively important for the
interpretation of experimental spectra, these effects provide an
additional degree of freedom for tailoring optical and transport
properties of quantum structures.

In this paper we examine two prototype examples relevant to the
physics of quantum wires (QWIs) and dots (QDs). We first consider
properly designed hybrid semiconductor/insulator QWIs based on GaAs,
and show that dielectric confinement (DC) may lead to excitonic states
with a binding energy exceeding the room temperature thermal energy
$kT_{\mbox{\tiny room}}$ (a prerequisite for exploiting excitonic
states in electro-optical devices) without degrading the optical
efficiency typical of conventional GaAs/AlAs nanostructures. Secondly,
we show that the dielectric constant of the environment may strongly
affect the addition spectra of QDs by modifying the electronic ground
state with respect to the case of good dielectric matching.

Our theoretical scheme \cite{Goldoni} moves from the following basic
considerations. When the dielectric constant $\epsilon({\bf r})$ is
spatially modulated, the Coulomb interaction between, say, two electrons
sitting at positions ${\bf r}$ and ${\bf r}'$, is given by $V({\bf
r},{\bf r}')= e^2 G({\bf r},{\bf r}')$, where $G({\bf r},{\bf r}')$ is
the Green's function of the Poisson operator, i.e.,
\begin{equation}
\bbox{\nabla}_{\bf r}\cdot\epsilon({\bf r}) \bbox{\nabla}_{\bf r}
G({\bf r},{\bf r}')= -\delta({\bf r}-{\bf r}').
\label{Poisson}
\end{equation}
Therefore, the space dependence of $\epsilon({\bf r})$ modifies 
$G({\bf r},{\bf r}')$ with respect to the homogeneous case, where 
$\epsilon({\bf r})=\epsilon_\circ$ and $G_\circ({\bf r},{\bf r}') = 
1 / [4\pi\epsilon_\circ |{\bf r}-{\bf r}'|]$.
This, in turn, modifies the Coulomb matrix elements between the 
quantum states of the structure\cite{SE,Goldoni} which, in the
basis ensuing from the single-particle envelope functions 
$\Phi^{e(e')}$,  can be written as
\begin{equation}
V_{ij} = e^2 \!\! \int \! \Phi_{i}^{e*}({\bf r})
\Phi_{j}^{e'*}({\bf r}')
G({\bf r},{\bf r}') \Phi_{i}^{e'}({\bf r}') \Phi_{j}^e({\bf r})
d{\bf r} d{\bf r}'.
\label{CME}
\end{equation}
Here $i$, $j$ stand for an appropriate set of quantum numbers
labelling the states.
If the symmetry of the structure is low, as in the realistic quantum 
wire structures considered below, Eq.\ (\ref{Poisson}) must be 
explicitly solved, and the ensuing potential is then used in (\ref{CME}). 
In this case it is convenient to cast Eqs.\ (\ref{Poisson}) and (\ref{CME}) 
in Fourier space as described in Ref. \onlinecite{Goldoni}. In contrast,
for particularly simple structures the analytic form of the potential
can be directly obtained in real space. This is, e.g., the case for the
second prototype structure discussed below, the spherical QD: here
two electrons can be shown to interact via the potential \cite{Brus}
\begin{equation}
V({\bf r}_i,{\bf r}_j)={e^2 \over \epsilon_1} {1\over
|r_i-r_j|}
+{e^2\over\epsilon_1R_d}
\sum_{k=0}^{\infty} {(k+1)(\epsilon-1)\over(k\epsilon+k+1)}
\left({r_i r_j\over R_d^2}\right)^{k}
P_k(\cos\Theta_{ij}),
\label{QDP}
\end{equation}
where $R_d$ is the QD radius, $\epsilon=\epsilon_1/\epsilon_2$, and
$\epsilon_1$ ($\epsilon_2$) is the dieletric constant of the inner
(outer) material. In the following we discuss the basic results 
and the relevance of these effects for our prototype QWIs and QDs.

{\bf (i) QWIs with remote dielectric confinement.}

\noindent
Recently, we have proposed that remote dielectric confinement (RDC)
\cite{Goldoni} may be used in order to enhance the exciton binding
energies $E_b$. In convential nanostructures $E_b$ is considerably
enhanced by quantum confinement; however, for GaAs-based structures,
observed values of $E_b$ are still well below $kT_{\mbox{\tiny room}}$
\cite{universal}. On the other hand, owing to the low optical quality
of typical semiconductor/oxide interfaces, oxides cannot be used
directly as confining barriers. Our novel approach is based on the
idea that {\em quantum and dielectric confinement can be spatially
separated since they are effective over different length scales}. In
the proposed structures the enhanced electron-hole overlap induced by
quantum confinement in conventional GaAs/AlGaAs structures is combined
with the DC provided by polarization charges which form at a {\em
remote interface} with a low-dielectric constant material, typically
an insulator; since electron and hole wavefunctions decay
exponentially into the barrier, they will not be affected by the
presumably disordered remote interface.

As an example of our approach, we discuss quantitative predictions for the case of a
conventional V-shaped GaAs/AlAs QWI with two oxide layers added above
and below the QWI at a distance $L$. The cross-section is shown in 
Fig.\ 1(a). The additional layers are characterized by a small dielectric 
constant that we take equal to 2 (see, e.g., Ref.\ \onlinecite{Fiore}). 
For this structure, we find $E_b = 29.3$ meV, to be compared with 
$E_b = 13\,\mbox{meV}$ of the conventional (i.e., with no oxide layers) 
structure. Fig.\ 1(a) 
shows that the origin of this dramatic enhancement is the large
polarization of the AlAs/oxide interfaces induced by the excited
electron and hole charge densities. A small polarization charge is
also induced at the GaAs/AlAs interface, due to the small dielectric
mismatch. In Fig.\ 1(b) we show the calculated $E_b$ for selected
values of $L$. Obviously, $E_b$ is maximum when the oxide layer is at
minimum distance \cite{limit}, $L=0$, where it is enhanced by more than 
a factor 3 with respect to $E_0$, and it is well above 
$kT_{\mbox{\tiny room}}$. The important point here is that
$E_b$ decreases slowly, indeed as $L^{-1}$, with the distance $L$, and
crosses $kT_{\mbox{\tiny room}}$ at $L$ as large as $\sim
9\,\mbox{nm}$, where the effects of the disorder at the Oxide/AlAs
interface are very small.

\begin{figure}[t]
\noindent
\unitlength1mm
\begin{picture}(180,50) 
\put(0,0){\psfig{figure=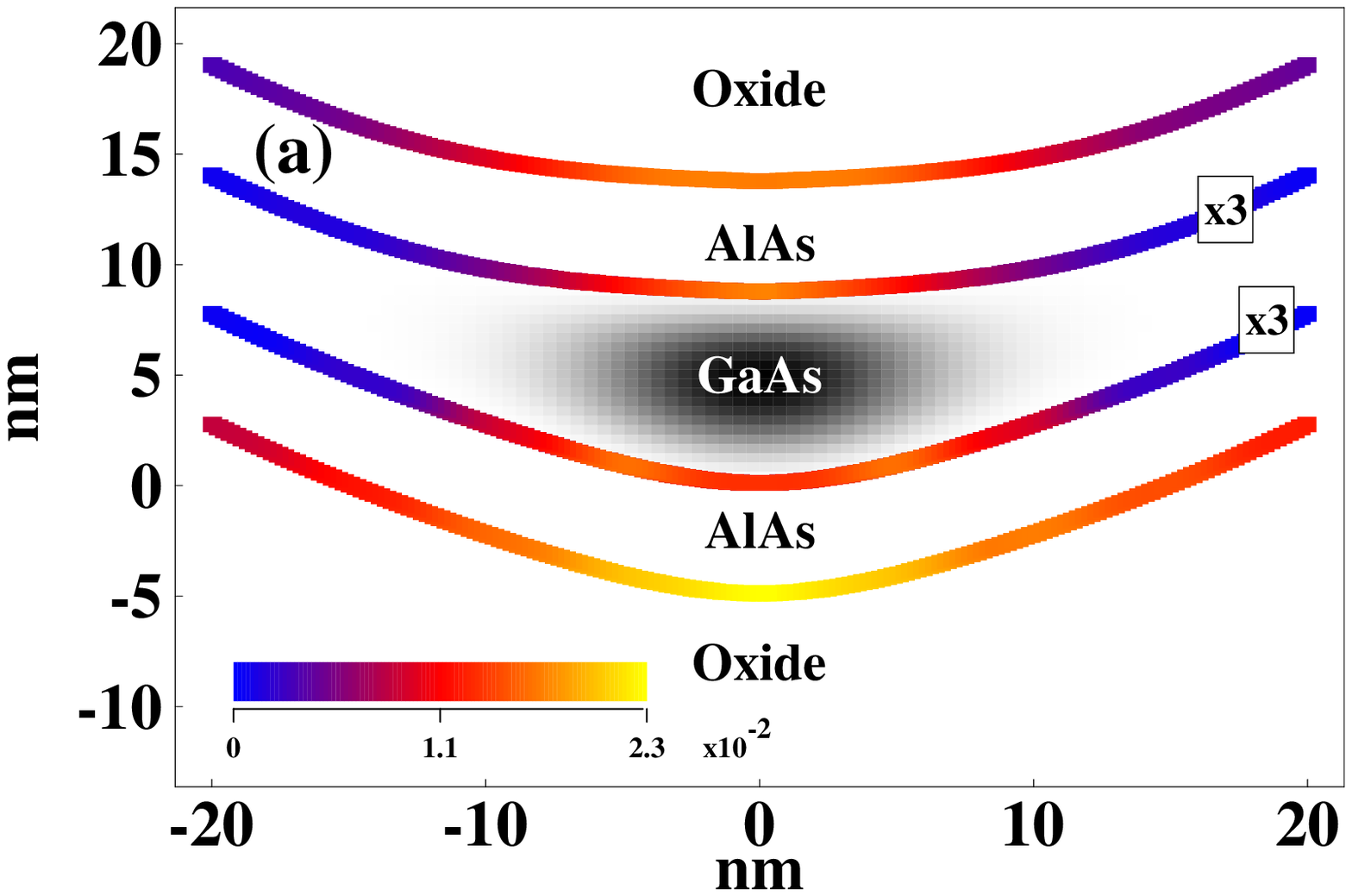,width=80mm}}
\put(82,20){\psfig{figure=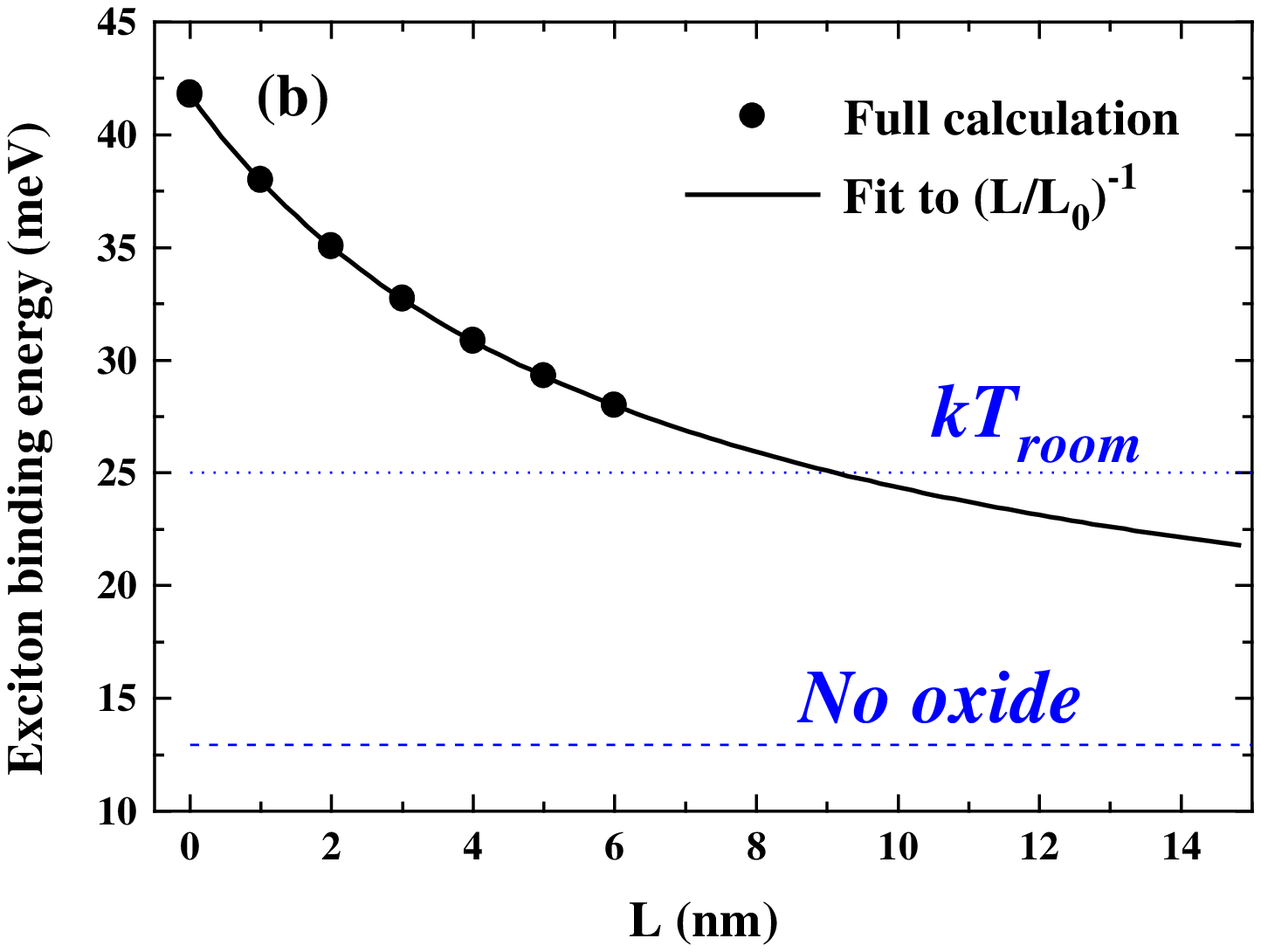,width=90mm}}
\end{picture}
\caption{(a) Cross section of a hybrid V-shaped QWI with the
interface polarization charge (units of nm$^{-1}$) and the
lowest-subband hole charge density (arbitrary units). The oxide layers
are at $L=5\,\mbox{nm}$ from the inner GaAs/AlAs interfaces. The
polarization charge at the GaAs/AlAs interfaces is multiplied by 3 for
clarity. (b) $E_b$ versus distance of the oxide layers from the inner 
interfaces. Dots: full calculation. The solid line is a fit to $L^{-1}$. 
Dashed line: binding energy $E_0$ of the corresponding conventional 
structure (no oxide layers).
Dotted line: thermal energy at $T_{\mbox{\tiny room}}=300\,\mbox{K}$.}
\end{figure}

{\bf (ii) Quantum dots in dielectric environments.}

\noindent
These structures have become accessible by transport studies only
very recently. \cite{Kle} They are III-V or II-VI nanoparticles
embedded in materials with different dielectric properties, such as organic 
matrices in a colloid. QDs in biological environments are also 
assuming increasing importance. The addition energies $E_{add}(N)$ 
(the energy required to add an electron to a QD containing
N electrons) have been used used to characterize these systems 
experimentally, but a theoretical description is still lacking. 
To obtain it, we must compute the ground state energy, $E_0(N)$, 
of the QD with $N$ interacting electrons (assumed to be confined in 
a spherical parabolic potential). 
The chemical potential of the QD with $N$ electrons is then 
$\mu(N)=E_0(N)-E_0(N-1)$,
from which we obtain 
\begin{eqnarray}
E_{add}(N)=\mu(N+1)-\mu(N)
\nonumber
\end{eqnarray}
The ground state energies $E(N)$, obtained from an Hubbard-like 
approximation to the many-body hamiltonian \cite{Rontani}, 
give rise to the addition spectra of Fig.\ 2. For $N \leq 5$, we have also 
performed the exact diagonalizations of the many-electron hamiltonian, 
with results that are almost identical.

The solid line in Fig.\ 2 is the calculated addition spectrum for a
dielectrically homogeneuos QD, i.e., $\epsilon=1$ in Eq.\ (\ref{QDP}).
The peaks at $N=2$ and $N=8$ correspond to the addition of one
electron to a QD with a closed s- and p-shell, respectively; the
weaker peaks at $N=5$ and $N=13$ correspond to the addition of one
electron to a QD with a half-filled outer shell where all spins are
parallel, as expected by a filling of the shells according to Hund's
rule. When $\epsilon>1$ the addition spectra of Fig.\ 2 are affected
in several ways. Let us first consider the behaviour for $N\leq8$. As
$\epsilon$ is increased, the spectra are shifted to higher energies,
since a larger energy is needed to add new electrons to the QD due to
the enhanced Coulomb repulsion. Note, however, that this shift is not
rigid, and the half-shell peak at $N=5$ is enhanced with respect to
full-shell peaks at $N=2$ and $N=8$. This result is quite general and
derives from the different combinations of direct and exchange Coulomb
terms that enter the ground state energies determining the full and
half-shell peaks.

The changes taking place at larger $N$ are more dramatic. As
$\epsilon$ is increased, the ordering and amplitude of the peaks
deviates from the behaviour at $\epsilon=1$. As can be seen Fig.\ 2,
half-shell peaks become comparable with full-shell ones, and
additional features appear for the larger values of $N$. Inspection of
the ground state configurations shows that this is due to a shell
filling in violation of Hund's rule. Above a critical value $N_c$, a
{\em reconstruction} of the electronic configuration takes place,
i.e., the added electron will not be arranged in the most external
shell, leaving the remaining electrons in the previous configuration.
Instead, it will cause other electrons in the inner shells to be
promoted to shells of higher angular momentum. This reconstruction
occuring at large values of the dielectric mismatch is similar to the
one predicted in QDs in a strong magnetic field (where a similar
enhancement of the Coulomb interaction takes place).

In summary, we have shown that dielectric confinement effects may
strongly affect quantum states in dielectrically modulated
nanostructures. By modulating the dielectric mismatch between
different layers it is possible to tune the Coulomb interaction
between the quantum confined states, in analogy to what is done by
external magnetic fields and/or doping, thereby modifying substantially 
the optical and addition spectra of nanostructures.

\begin{figure}[t]
\noindent
\unitlength1mm
\begin{picture}(180,50) 
\put(10,0){\psfig{figure=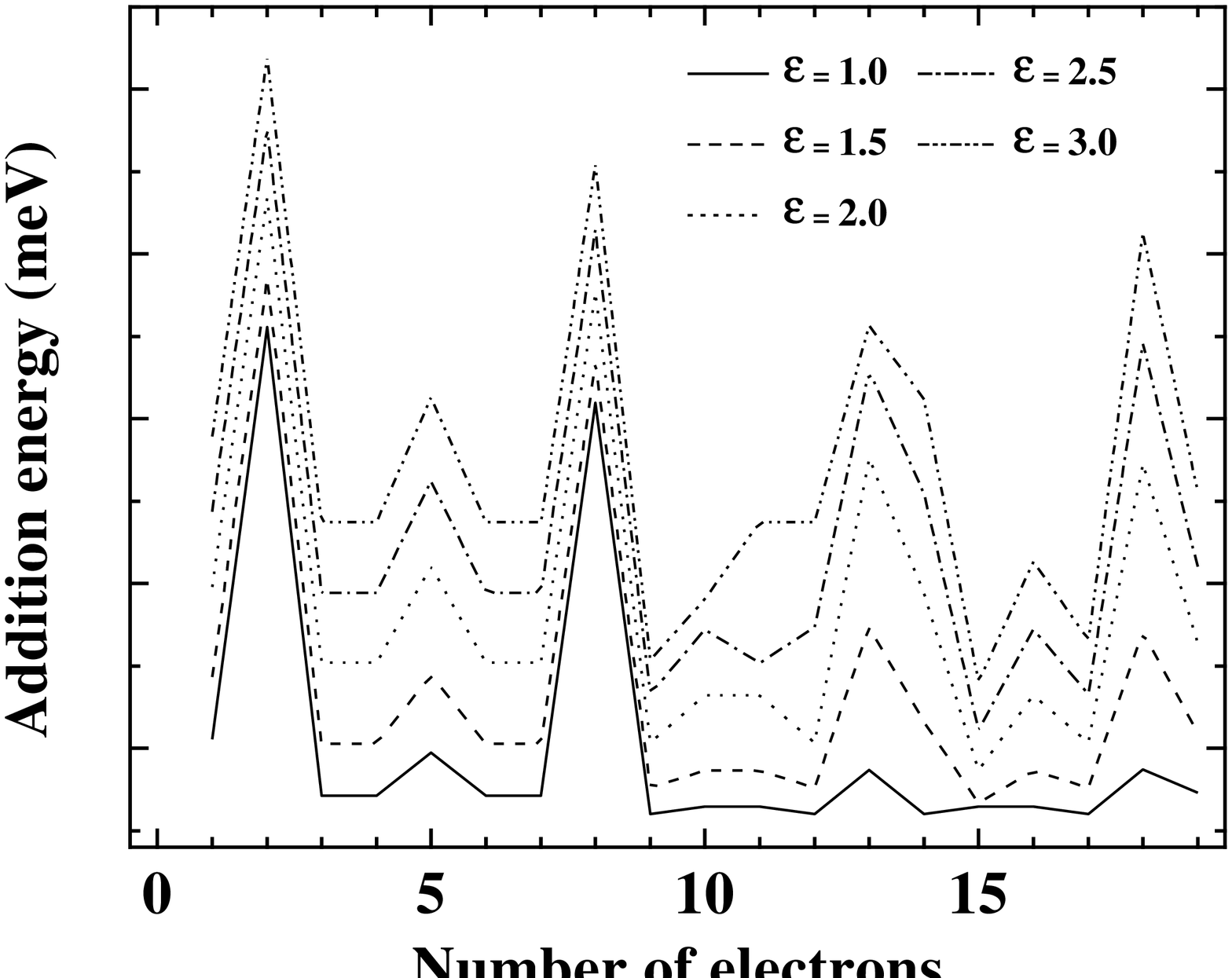,width=80mm}} 
\put(40,40){\lcaption{60truemm}{0truemm}{Addition spectra obtained
with the Hubbard approximation for several values of the dielectric
mismatch $\epsilon$. The parameters used in the computations are for a
CdSe \protect\cite{LaBo} spherical QD of radius of 3 nm, that corresponds to a
confinament energy of about 150 meV.}}
\end{picture}
\end{figure}

\vspace{2truemm}
This paper was supported in part by MURST-40\% through grant "Physics
of nanostructures" and by INFM through grant PRA-SSQI.

%



\begin{references}
\bibitem{SE} The renormalization of the interaction gives also rise to a
self-energy term; in most cases this term is small and will not be
discussed in this paper. See also Ref.\ \cite{Goldoni}.
\bibitem{limit} Our calculations have been extended
down to $L=0$ for completeness, although in this limit they should be taken
with caution, due to the neglect of the AlAs/oxide interface effects.
\end{references}
\end{document}